\documentclass[elsart,12pt]{article}
\usepackage{graphicx}
\begin{document}
\begin{center}
\Large {Cosmic Rays and the Monogem Supernova Remnant}\\
\end{center}
\vspace {0.5cm}  
\begin{center}
\footnote{E-mail address: erlykin@sci.lebedev.ru} 
A.D.Erlykin~$^{(1,2)}$, A.W.Wolfendale$^{(2)}$
 
\end{center}
\begin{flushleft}
(1) {\em P. N. Lebedev Physical Institute, Moscow, Russia}\\
(2) {\em Department of Physics,University of Durham, Durham, UK}\\
\end{flushleft}

\begin{abstract}
Recent findings indicate that the Monogem Ring and the associated pulsar 
PSR B0656+14 may be the `Single Source' responsible for the formation
of the sharp knee in the cosmic ray energy spectrum at $\sim$3PeV. The
energy spectum of cosmic rays 
expected for the Monogem Ring supernova remnant (~SNR~) from our SNR 
acceleration model \cite{EW9} has been published by us elsewhere
\cite{EW7}. In this paper we go on to estimate the
contribution of the pulsar B0656+14 to the cosmic rays in the PeV
region. We conclude that although the pulsar can contribute to the formation
of the knee, it cannot be the dominant source of it and an SNR is still needed. 

We also examine the possibility of the pulsar giving
the peak of the extensive air shower (~EAS~) intensity observed from
the region inside the Monogem Ring \cite{Chil1}. The estimates of the
gamma-ray flux produced by cosmic ray particles from this pulsar
indicate that it can be the source of the observed peak, if the
particles were confined within the SNR during a considerable fraction of its
total age. The flux of gamma quanta at PeV energies has a high
sensitivity to the duration of the confinement. The estimates of this 
time and of the following diffusion of cosmic rays from the confinement
volume turn out to be
in remarkable agreement with the time needed for these cosmic rays to
propagate to the solar system and to form the observed knee in the
cosmic ray energy spectrum. 

Other possible mechanisms for the production of particles which could
give rise to the observed narrow peak in the EAS intensity were also
examined. Electrons scattered on the microwave
background or on X-rays, emitted by SNR, can not be responsible for
the gamma-quanta in the peak. 
Neutrons produced in PP - collisions or released from the
disintegration of accelerated nuclei seem to be also unable to create
the peak since they cannot give the observed flux.  

    If the experimental EAS results concerning a point-like source are 
confirmed, they can be important, since\\
(i) they will give evidence for the acceleration of protons
or heavier nuclei by the pulsar; \\
(ii) they will give evidence for the existence of a confinement mechanism in SNR
;\\
(iii) they will confirm that cosmic rays produced by the Monogem Ring SNR
and associated pulsar B0656+14 were released recently giving rise to
the formation of the sharp knee and the observed narrow peak in the
EAS intensity~;\\
(iv) they will give strong support for the Monogem Ring SNR and the
associated pulsar B0656+14 being
identified as the Single Source proposed in our Single Source Model
of the knee.

A number of predictions of the examined mechanism are made.

\end{abstract}
\section{Introduction}

A few years ago we suggested the `Single Source Model' to explain the remarkable 
sharpness of the knee in the cosmic ray energy spectrum at $\sim$3 PeV \cite{EW1,EW2}, 
a feature noticed even in the first publication on this subject, 46
years ago \cite{Kulik}. The model is based
on the assumption that a single, 
relatively recent and nearby supernova remnant contributes significantly 
to the cosmic ray intensity at PeV energies. The sharpness is due to the cutoff
in the energy spectrum of cosmic rays accelerated by SNR. According to the theoretical
model of the SNR acceleration mechanism developed by Berezhko et al. \cite{Berez}
the cutoff for the acceleration in the hot and low density interstellar medium (~ISM~)
is at a rigidity of $\sim$0.4 PV. To match the position of the knee at an energy of
$\sim$3 PeV and to explain the second peak of the intensity at $\sim$10-15 PeV, 
observed in most of the experiments, the model assumes that the Single Source emits 
predominantly medium (~oxygen~) and heavy (~iron~) nuclei with an admixture of 
sub-iron nuclei. This assumption is reasonable, in view of the ISM,
which provides the nuclei, having been seeded by a previous supernova 
(~see the next paragraph~). Comparing the shape of the energy spectrum of cosmic rays from the 
Single Source and its total energy content with the model of SNR acceleration and 
the propagation of cosmic rays through the ISM we derived a likely interval of
distance (~230-350 pc~) and age (~84-100 kyear~) for the Single Source \cite{EW3}. 

On the basis of our estimates of distance and age we calculated the possible flux of high energy 
gamma rays from the Single Source and found that it is unlikely to be observed  
at sub-GeV
and TeV gamma rays with gamma telescopes of the present sensitivity \cite{EW7}. 
The reason is that being nearby the Single Source is in our Local Superbubble with its 
low (~$\sim 3\cdot 10^{-3} cm^{-3}$~) density of target gas and the SNR
should not be a discrete source , but extended with an angular radius of 
$\sim$20$^\circ$. It is difficult to detect an excess intensity from
such an extended source
since the estimates of the background are very unreliable. Among the sources 
which would satisfy these limits of distance and age we indicated the Monogem Ring 
and Loop I \cite{EW4}. The same sources were also discussed in connection with their 
possible contribution to the flux of high energy electrons 
\cite{Koba1,EW5,Koba2}  

Recently, Thorsett et al. \cite{Thors}, using the triangulation technique found the distance of the 
pulsar PSR 0656+14 associated with the SNR Monogem Ring. It is 288$\pm$30 pc and its 
spin-down age is $\sim$110 kyears, both of which are in remarkable agreement with our 
estimates for the 
Single Source \cite{EW7}. Such determinations had not previously been
possible from observations of the SNR itself. Thorsett et al. themselves claimed that the SNR Monogem Ring and its associated 
pulsar PSR 0656+14 can be the Single Source responsible for the
formation of the knee.

Armenian physicists have studied the sky near the Monogem Ring in the
sub-PeV and PeV range using the EAS technique and found a 6$\sigma$ excess
of the EAS intensity in one of their angular bins \cite{Chil1}; we
refer to this interesting result as the `Armenian peak'. Since their bin
of $3^\circ \times 3^\circ$ is narrower than the size of the SNR it is
thought that the excess is not due to the extended source, but to
a discrete source, viz. the pulsar.
     
Bhadra \cite{Bhadr} analysed theoretically the possibility for a
pulsar to be the Single Source, and concluded 
that the most likely pulsar candidates are Geminga and Vela.

In this paper we analyse the possibility of the pulsar PSR B0656+14, 
associated with 
the SNR Monogem Ring, being the Single Source responsible for the knee
and also see to what extent the `Armenian peak' could come from the
same object. Independently, we search for other evidence which might
confirm the reasonableness of the peak.

\section{Calculation of the pulsar energy spectrum}  

At the beginning we consider the pulsar as an isolated neutron star.  
The difference between the temporal dependence of particle acceleration by such a 
pulsar and by the SNR is that in the former case the process is continuous from the 
very beginning. At any time instant three processes follow, sequentially: emission of 
accelerated particles by the pulsar, propagation through the 
ISM and leakage from the Galaxy. On the contrary the acceleration by the expanding SNR shock wave is most likely 
not accompanied by immediate particle emission, rather, due to the compression of 
magnetic fields, the particles are confined within the shell for some time after the SN 
explosion and then released to `outer space' followed by diffusion
and eventual escape from the Galaxy. For an isolated pulsar, however, we
postulate that the other radiations, associated with it, so perturb the
ambient magnetic field that the very energetic particles have easy,
and prompt egress. 

The energy spectrum of emitted particles for an isolated pulsar was
considered as being equal to a mean spectrum emitted by a number of pulsars 
and averaged over an isotropic distribution of $\theta$, where $\theta$ is an 
angle between the spin axis and magnetic dipole axis of the pulsar.
The spectrum has been deduced on the basis of works \cite{Ostri,Blasi,Gille} in the
framework of the following scenario. If the emitted particles
are beamed along the magnetic dipole axis, then
their total number will be proportional to the solid angle as
$\frac{dN}{d\theta} \propto sin\theta$. An energy spectrum
of accelerated particles is connected with $\frac{dN}{d\theta} $ as 
$\frac{dN}{dE}=\frac{dN}{d\theta} / \frac{dE}{d\theta} $. If the particle
energy $E$ changes with $\theta$ as $E = E_{max} sin\theta $ then 
$\frac{dE}{d\theta} = E_{max} cos\theta $ and $\frac{dN}{dE} \propto
tan\theta = \frac{E/E_{max}}{\sqrt{1-(E/E_{max})^2}}$. We applied this averaged spectrum for the
case of an individual pulsar, because of the lack of knowledge of 
the particular value of $\theta$.

We adopted the total energy contained in this spectrum as being equal to
$E_{max}$. The further normalization to the total rotation energy loss
$\dot{E_{rot}}$ results in the particle emission rate as a function
of their energy and time being given by:
\begin{equation}
\frac{d^2N}{dEdt}=\frac{4\dot{E_{rot}}E}{\pi E_{max}^3 \sqrt{1-(E/E_{max})^2}}
\end{equation}   
Here {\em E} is the particle energy, {\em t} is the time since the
creation of the pulsar,
$E_{max}$ is the maximum energy of the emitted particles, which is equal to
\begin{equation}
E_{max} = \frac{E^0_{max}}{1+2t/T_0}
\end{equation}
where $E^0_{max}(GeV) = 3\cdot 10^{-7} Z \sqrt{\frac{3I\dot{P}_0}{2cP_0^3}}$
is the maximum energy at $t=0$, {\em Z} is the particle charge, {\em I} - 
the pulsar moment of inertia, taken as
10$^{45}$ gcm$^2$, $P_0, sec$ and $\dot{P}_0$ - the initial period of
rotation and its time derivative respectively, {\em c} - the speed of
light and $T_0 = P_0/\dot{P}_0$.  
$\dot{E_{rot}}$ is the rate of the rotation energy loss, which is equal to
\begin{equation}
\dot{E_{rot}}=\frac{\dot{E}_0}{(1+2t/T_0)^2}
\end{equation}
where $\dot{E}_0$ is the rate of energy loss at the initial time, $t=0$.
Here we assume that all the rotational energy lost by the pulsar is
given to cosmic ray 
particles; this is certainly an extreme assumption, but it gives an upper limit for 
our estimates of the cosmic ray intensity.

There is a minor difference between our consideration and that of 
the authors of \cite{Ostri,Blasi,Gille}. They assumed that at any
instant the pulsar emits monoenergetic particles, i.e. their spectrum
looks like a line at the energy $E_{max}$. Inspite of this difference,
our expression (1), with its evident divergence in the denominator,
which follows from the mathematical treatment of the adopted
assumptions and not necessarily is realized in nature, when
integrated over the pulsar lifetime gives the same well known
spectrum $\frac{dN}{dE} \sim E^{-1}$, as in the case of monoenergetic 
particles.    
   
The propagation model includes both diffusion and escape from the
Galaxy. Such a combination is used, because though the diffusion
has been considered in the non-uniform ISM this non-uniformity did not
include effects connected with the finite dimensions of the Galactic
disk and the ensuing escape of particles from the process of diffusion
and from the Galaxy. 

The survival against leakage from the Galaxy is described by the
usual expression, taken from the leaky box model:
\begin{equation}
S(t,T,E)=exp(-\frac{T-t}{\tau(E)})
\end{equation}
where $T$ is the spin-down age of the pulsar, $\tau(E)$ is the mean life time of particles
in the Galaxy, which depends on the particle energy $E$ and charge $Z$ as 
$\tau(E), year = \tau_0 (E/Z)^{-\delta}$, with $E$ - in GeV, $\tau_0 = 4\cdot 10^7$ year
and $\delta = 0.5$ in our part of the Galaxy \cite{EW6}.  

The diffusion of cosmic rays from the pulsar to the solar system is described by 
`anomalous diffusion' in the fractal-like ISM with the index $\alpha = 1$
\cite{Lagut,EW8,EW7}. The density of cosmic ray particles at a distance {\em R} from 
the source, emitted and observed at the time {\em t} and {\em T} respectively, is 
described for the case of spherical diffusion by
\begin{equation}
\rho(t,T,R,E) = \frac{1}{\pi^2R_d^3 (1+(R/R_d)^2)^2}
\end{equation} 
Here $R_d(E)=H_z (\frac{T-t}{\tau(E)})^{\frac{1}{\alpha}}$ is the
diffusion radius for our assumed scale height of $H_z$ = 1000 pc.
For comparison, we also considered the propagation, described by normal gaussian 
diffusion, with 
\begin{equation}
\rho(t,T,R,E) = \frac{1}{8\pi^{\frac{3}{2}}R_d^3}exp(-(R/2R_d)^2)
\end{equation}
and $\alpha = 2$ \cite{Lagut}.
 
The energy spectrum of cosmic rays is calculated as
\begin{equation}
\frac{dN}{dE}=\int_0^T \frac{c}{4\pi} \frac{d^2N}{dtdE} S(t,T,E) \rho(t,T,E,R) dt
\end{equation}
The observed parameters of the pulsar PSR 0656+14 are $P_0$ = 0.3848 s, 
$\dot{P} = 5.5032\cdot 10^{-14}$ \cite{Taylo}. We took the age of this pulsar as
$T = 1.005\cdot 10^5$ year, a value of 10 years for the $T_0$ parameter and
$R = 288$ pc for the distance \cite{Thors}. The calculations show that the spectrum
does not depend on the initial conditions for a wide interval of $T_0: 1 - 100$ 
years. For these numerical values of the parameters, $E_{max}^0 = 4.83\cdot 10^9$ GeV,
$\dot{E}_0 = 4.08\cdot 10^{53}$ GeV year$^{-1}$. Calculations have
been made for the energy spectrum
of cosmic ray protons (~{\em Z} =1~) and oxygen nuclei (~{\em Z} =
8~), (~assuming that oxygen nuclei could, in fact, be taken from the
pulsar surface and accelerated by it~). The result
is only weakly dependent on the mode of diffusion. The position of the peak in the energy 
spectrum is practically the same for both normal and anomalous diffusion. The energy  
contained in {\em observed} cosmic rays is, for the case of anomalous diffusion, higher by a factor
of 2.5, compared with that for the normal diffusion. The result for anomalous 
diffusion is shown in Figure 1. 
\begin{figure}[htb]
\begin{center}
\includegraphics[width=10cm,height=10cm,angle=0]{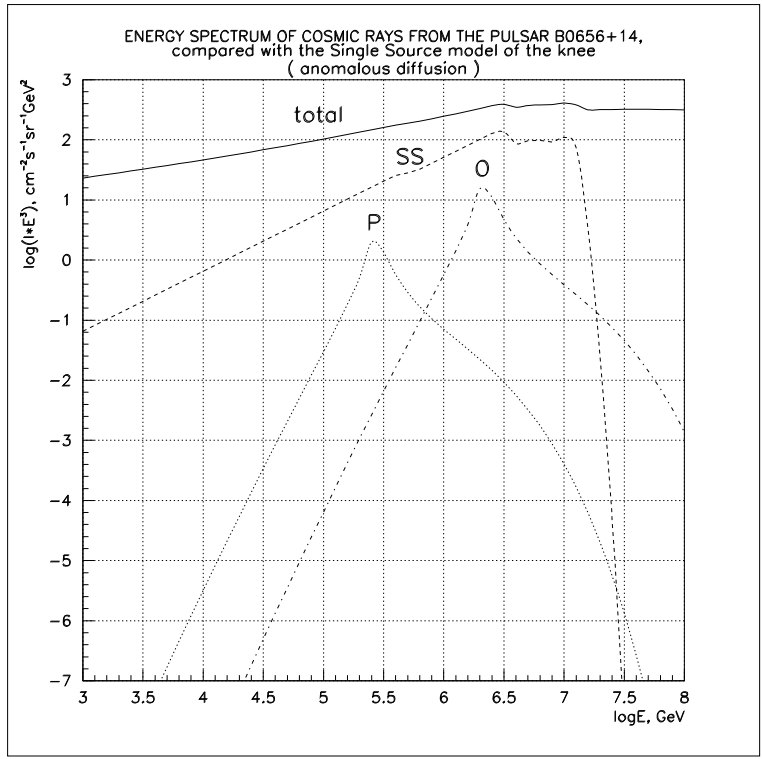}
\caption{\footnotesize The energy spectrum of cosmic rays from PSR B0656+14 observed 
at the present time and compared with the Single Source model of the
knee \cite{EW1,EW2}: full line - total energy spectrum of cosmic rays;
dashed line: energy spectrum of the Single Source, dotted line: energy
spectrum of protons accelerated by B0656+14, dash-dotted line - the
same spectrum for oxygen nuclei. The knee in the actual spectrum is at 
$logE(GeV) \approx 6.5$, close to our pulsar prediction for oxygen.}
\end{center}
\label{fig:puls1}
\end{figure}

\section{Contribution of an isolated pulsar to the knee}

It is remarkable that the spectrum of cosmic rays has a very 
sharp peak at a rigidity of 0.25 PV, i.e. a rigidity close to that of
the knee. It is the maximum rigidity of the particles 
emitted by the pulsar at the present time (~neglecting the time for
light to travel~).
The rapid rise of the spectrum below the peak is due to the shape of the 
emitted spectrum, which is $\propto E$ (~up to a maximum~) at any time instant during the whole life of the 
pulsar (~see (1)~). The leakage from the Galaxy and the anomalous diffusion through 
the ISM do not distort this dependence very much. 

The steep drop of the spectrum beyond the peak is due to three
reasons. The particles of these high energies can be emitted only
during a fraction of the pulsar's life time: the higher is the energy
- the smaller is that fraction. The escape from the Galaxy
is also high and increases with energy. The diffusion of the
particles is also very rapid and at the present time almost all the high
energy particles have passed the solar system and vanished into space.
The diffusive nature of the particle propagation means that the
spatial distribution of the particles will be nearly isotropic at the Earth.

The formation of the pulsar spectrum is illustrated in Figure 2, where the particle 
emission rate, their density after diffusion and the survival probability
against escape from the Galaxy are shown at different moments of the pulsar's 
history.
\begin{figure}[htbp]
\begin{center}
\includegraphics[width=15cm,height=15cm,angle=0]{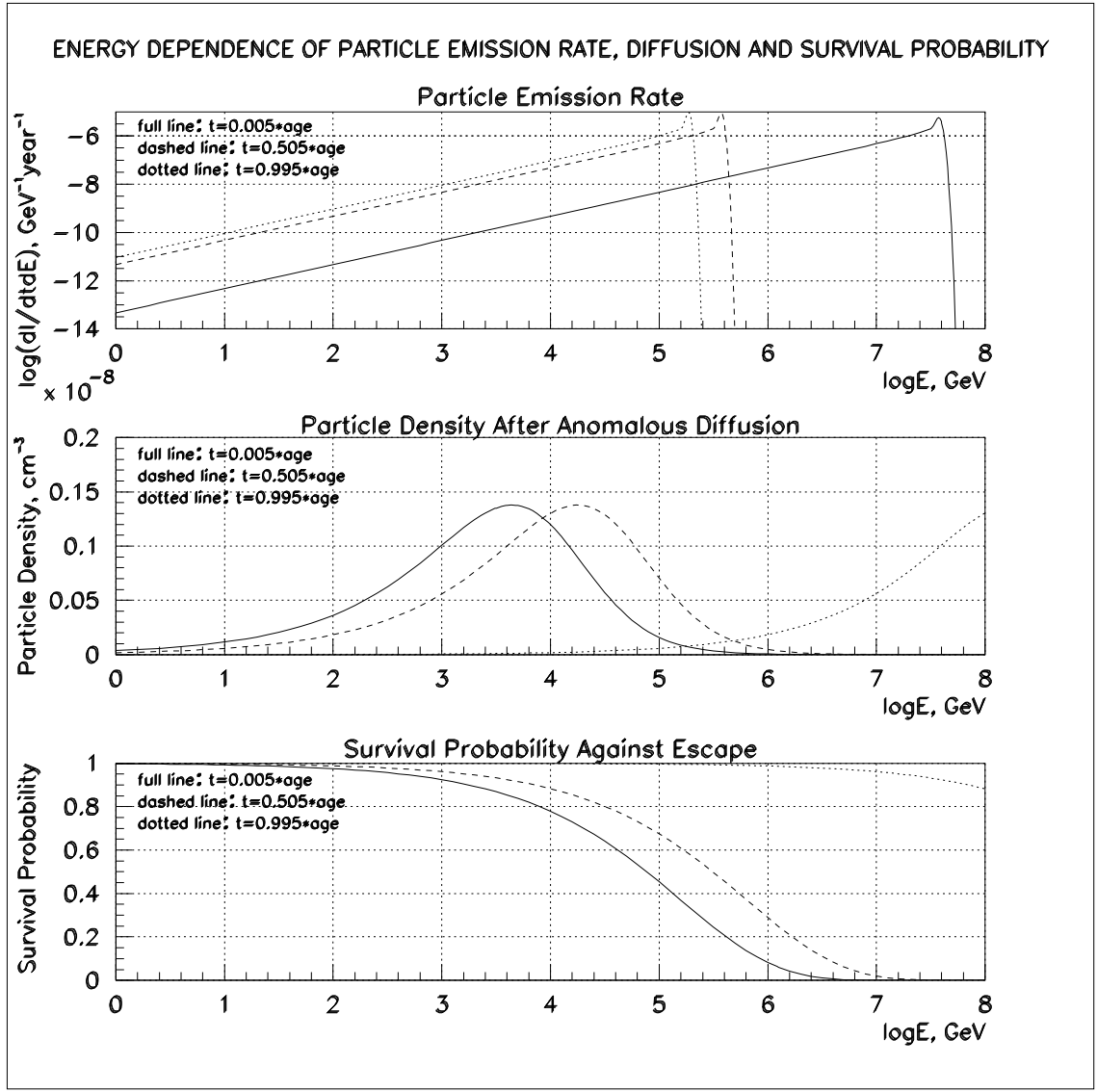}
\caption{\footnotesize The formation of the energy spectrum of
particles from the pulsar at different 
instants of its history: 0.005 (~full lines~), 0.505 (~dashed lines~) and 0.995 
(~dotted lines~) of its age. The upper graph: the 
particle emission rate, the middle graph: the particle density at
Earth after anomalous 
diffusion and the lower graph: the survival probability against escape from the 
Galaxy.}
\end{center}
\label{fig:puls2}
\end{figure}

The energy density contained in the proton and oxygen spectra is the same and equal to
$1.9\cdot 10^{-6} eVcm^{-3}$. Despite the fact that we 
normalized the energy transferred to the cosmic rays to the total loss of the rotation 
energy the cosmic ray energy density turns out to be small compared with 
the value of $\sim 2.24\cdot 10^{-4} eVcm^{-3}$ needed to form the knee in the SS model
\cite{EW7}. If, instead of energy density, we compare the intensity at the knee needed to ensure the observed 
cosmic ray flux and intensity in the peak for Z = 8, the difference becomes 
smaller, but it is still rather large (~$\simeq
7$~). We conclude, therefore, that 
the pulsar PSR 0656+14, if it is considered as an isolated neutron
star, can contribute up to $\sim15$\%
to the formation of the knee, due to the sharpness of its energy
spectrum and the 
closeness of its peak rigidity to the needed value of 0.4 PV, but it
seems not to be able to produce enough cosmic rays to be the dominant
source of the knee.

\section{The EAS intensity peak in the Monogem Ring region and the
possibility of associating it with the pulsar B0656+14}

\subsection{Observation of the peak}

It has been mentioned in the Introduction that since the Monogem Ring SNR is located in our Local Superbubble, with
its low gas density,
 and it is not discrete, but an extended source, which occupies a substantial
part of the sky with an angular size of about 25$^\circ$, we do not expect a 
measurable flux of high energy gamma quanta from it
\cite{EW7}. However, Armenian 
physicists looking for regions with an excessive flux of EAS at PeV
energies have found such a domain within the Monogem Ring SNR \cite{Chil1}
(~the `Armenian peak'~).
Their search bin had a size $3^\circ \times 3^\circ$, which is not point-like, but 
definitely smaller than the size of the Monogem Ring SNR itself. 
The magnitude of the excess was about 6 standard deviations and
 therefore appears well founded statistically.  

The immediate idea, to be examined now, is that inside such an
extended source as the Monogem Ring SNR 
there is an additional discrete source of high energy cosmic rays,
which gives this excess. The most plausible discrete source within the SNR is the pulsar, specifically PSR B0656+14. Though the position of 
the peak is displaced from the present pulsar position, it is
reasonable to analyse the probability of this pulsar producing the 
observable peak. The inevitable diffusive scattering of
particles from the SNR and pulsar means that the Armenian peak must be due to
gamma rays or neutrons. 
     
An intensity feature of the excess flux is that its spectrum appears
somewhat flatter than that of the background flux; such a feature adds
to its veracity.

\subsection{The observed flux and the energy of particles
responsible for the peak}

Chilingarian et al. \cite{Chil1} do not give an estimate of the observed excessive 
flux of EAS. 
However, it is possible to make such an estimate on the basis of their published 
results. The most reliable peak is established for EAS with a total size 
$N_e > 10^6$. The intensity of such EAS in the vertical direction at
the Aragats altitude
of 3200 m above sea level, 
where the MAKET ANI array is operating, is $(3.3\pm 0.7)\cdot 10^{-11}$
cm$^{-2}$s$^{-1}$sr$^{-1}$ \cite{Adamo,Chil2}. The mean number of counts per bin in 
the studied declination band (~12.5$^\circ < \delta < $ 15.5$^\circ$~) is 18.46.
To get this number of counts one needs to have an exposure 
$S\Omega T = (5.6\pm 1.2)\cdot 10^{11} cm^2\cdot s \cdot sr$. The solid angle for
the $3^\circ \times 3^\circ$ bin is $2.74\cdot 10^{-3}$ sr, hence 
$ST = (2.04\pm 0.44)\cdot 10^{14} cm^2s$. The excess number of counts in the 
discovered bin was 25$\pm$6, and in order to get this number one needs to have a flux of $(1.2\pm0.4)\cdot 10^{-13} cm^{-2}s^{-1}$.      

If these showers are produced by gamma quanta, their energy for the shower size 
$N_e > 10^6$ at 3200 m above sea level should be more than 1.07 PeV
\cite{Plyas}. If the showers are produced by neutrons, their energy
must be higher and for $N_e > 10^6$ at Aragats level should exceed 2.5
PeV \cite{Knapp}.

\subsection{Particles from an isolated pulsar}

Since the pulsar B0656+14 is at a distance of about 300 pc from the
solar system, then if the observed gamma quanta or neutrons are
produced by protons from it, they can be born only 900 
years ago, i.e. at the present epoch. From the results presented in \S2 it
is clear that 
the pulsar B0656+14, if it is an isolated neutron star (~i.e. the
particles can diffuse freely from it~), cannot give particles 
above its peak energy of 0.25 PeV at the present epoch. Higher energy 
particles produced in the past would have the necessary higher energy
but they have already diffused for a long time and their density in the 
vicinity of the pulsar at the present epoch is very low. Higher energy 
particles were also produced in the past, but the majority have
already passed beyond the solar system. Heavier nuclei, if they are 
accelerated by this pulsar, and have higher total energy at the
present time, cannot help, since they have even smaller energy per
nucleon in the peak of their energy spectrum (~Figure 1~).

Therefore, if the pulsar B0656+14 is isolated it cannot easily give
the Armenian peak at energies above 1 PeV. The word `easily' is used
because it is just possible that the effective magnetic field is
higher than adopted and with it, $E_{max}$. This possibility arises
because of the fact pointed out in \cite{Phinn}, and used by us in
connection with the origin of very energetic particles \cite{EW10}, that
the field may have a complicated topography, so that higher multipoles
of the field have bigger values than that of the dipole. Nevertheless,
the diffusion propagation during the pulsar age has to reduce
the cosmic ray density at high energies to such low values that they
would not be able to give a measurable effect.  

\subsection{Particles from the pulsar associated with a SNR}   

There is a way to include into consideration higher energy particles born 
in the past, but which, however, produce gamma quanta or neutrons at
the present time and this is to 
associate the pulsar with a SNR, since they were both born in the same
SN explosion and reject the assumption that the pulsar can be regarded
as isolated. In this way we allow the produced particle to be trapped in the SNR
in the usual manner for SNR-accelerated particles
\cite{EW9,Koba2,Voelk}. The pulsar created as a result of the 
explosion is located within the shell close to the SNR morphological
center. We now assume that cosmic rays accelerated by the pulsar are also confined 
for the same time as those from the SNR. They are {\em all} released much later, begin to 
diffuse from internal regions of the SNR and eventually escape from the Galaxy. 
Their density in the vicinity of the pulsar is still high. In the process of diffusion 
through the ISM they produce qamma quanta which can be seen
now. This is the scenario.        

In fact, the problem of particle escape will be complicated not least
because of the effect of the pulsar wind on the ambient ISM in, and
near, the ISM. In the original version of the 'Single Source Model' 
\cite{EW1,EW2}, where pulsar-effects are ignored, we assume that the
SNR-accelerated particles are {\em all} trapped until the average
cosmic ray energy density is equal to the value outside the remnant,
when they all escape. We continue with this approximation although if
shorter confinement times are allowed for higher energy particles the energy
spectrum is steeper. This effect should not be important when the
confinement time is much less than the pulsar and SNR age. In the case 
of Monogem SNR it is not completely true, but we adopt this assumption 
as a first approximation. 
  
    We calculate energy spectra of cosmic rays accelerated by the pulsar B0656+14
and observed now at the Earth assuming that the confinement time is the fraction
of the pulsar age. The results are shown in Figure 3.
\begin{figure}[htbp]
\begin{center}
\includegraphics[width=15cm,height=15cm]{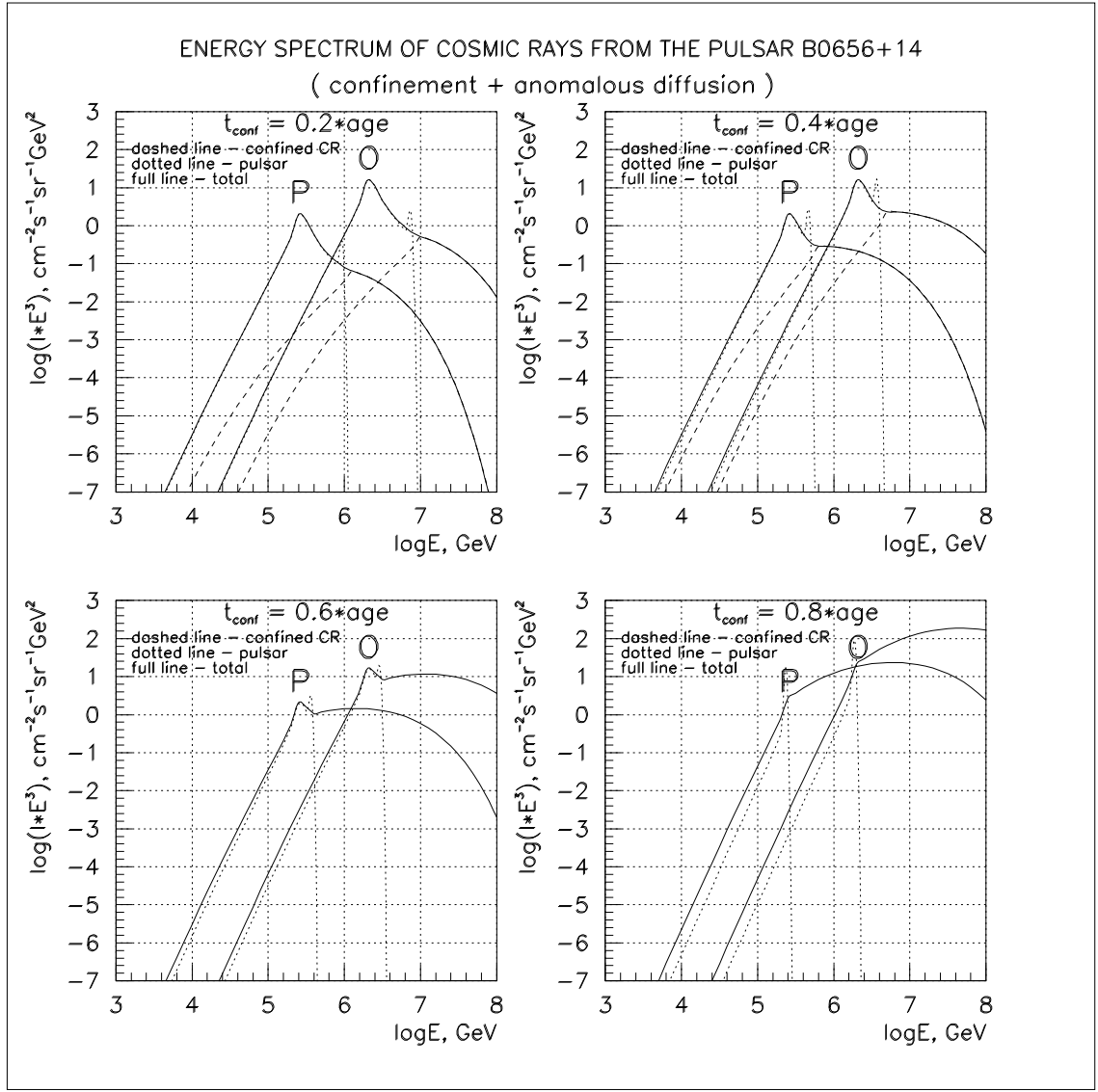}
\caption{\footnotesize The energy spectrum of cosmic rays from the
pulsar B0656+14 observed at the Earth, calculated for different
times $t_{conf}$, during which cosmic rays were confined within the
SNR shell and after that they were released, 
begin to diffuse through the ISM and escape from the Galaxy. Dashed lines indicate the 
contribution from the cosmic rays accumulated during the confinement time, dotted lines
show the contribution from the pulsar since the end of the confinement. Full lines show
the total spectrum composed of these two. The spectra are
 shown for the confinement time lasting (a) 0.2; (b) 0.4; (c) 0.6 and
(d) 0.8 times the pulsar age.}
\end{center}
\label{fig:puls3}
\end{figure}
It is seen that as the instant when cosmic rays are released
from confinement
 approaches the present moment, the more highest 
energy cosmic rays remain within the region between the pulsar and the Earth and
can be the potential source of observed gamma quanta. Comparison with
 Figure 1 shows the `value' of the SNR trapping: particles above the
 knee (~at 3 PeV~) have now a higher intensity and thus this model is
 better able to explain the observation of particles of these high
 energies in practice.

\subsection{Gamma rays from the pulsar associated with the SNR}

\subsubsection{Proton and nuclei interactions}

We have calculated the expected
flux of gamma quanta with energy above 1 PeV as the integral along the
line of sight for gamma quanta produced in PP-collisions of protons
accelerated by the pulsar with the hydrogen atoms of the ISM:
\begin{equation}
F(>1PeV) = \int_0^R 2 c dr\int_{1PeV}^{E_{max}^0} \frac{dN}{dE}
\rho_{cr} \sigma_{in} n_{\gamma}(>1PeV) \rho_{ISM} dE 
\end{equation}     
Here $\frac{dN}{dE}$ is the energy spectrum of cosmic rays emitted by
the pulsar, 
confined during the time $t_{conf} = \Delta * age$, then released and, until the present time, 
survived after escape and diffusion. These spectra, and some other
input parameters,
are shown in Figure 4. $\sigma_{in}$ is the inelastic cross-section of
PP collisions, 
which has been taken as $\sigma_{in} = 0.82*(38.5+0.5ln^2(S/137)), mb$
with $S = 2M_p(2M_p+E)$ as 
the squared energy of PP-collision in the center of mass system, $M_p
= 0.938 GeV$ 
being the mass of the proton. $n_{\gamma}(>1PeV)$ is the multiplicity
of gamma quanta with energy above 1 PeV. It was calculated using
the algorithm and formulae given in \cite{EW3} and shown in the lower
panel of Figure 4. $\rho_{ISM}$ is the {\em mean} density of the target gas in
the ISM, taken as $3\cdot 10^{-3} cm^{-3}$ since B0656+14 is situated
in our Local Superbubble with its low gas density. 

The term $\rho_{cr}$ describes the lateral distribution function
(~LDF~) of the cosmic ray density and is inversely proportional by the
volume $V_{cr}$ occupied by the cosmic rays. Here we follow the scenario 
described in \cite{EW3}. The particles emitted by the pulsar were
confined during
the time $t_{conf}$ within the expanding spherical SNR shell with
radius $R_{s} = 50 \sqrt{\frac{t_{conf}}{2\cdot 10^4 year}}$, pc. Inside
this shell they are completely isotropised and their lateral
distribution is uniform. After the confinement time they are released
and begin
to diffuse through the ISM. As before we have taken a spherical mode
of anomalous diffusion with no influence of beaming (~see (5)~), which
is a reasonable assumption. The LDF of cosmic ray density 
was calculated as $\rho_{cr} = \frac{1}{V_{cr}Q(E,r,t)}$ with
\begin{equation}
V_{cr}(E,R_s) = \frac{4}{3} \pi
(R_{s}^3+3R_d^3(0.7854+(\frac{R_s}{R_d})\cdot
0.9986+(\frac{R_s}{R_d})^2\cdot 0.7849))
\end{equation}   
where the diffusion radius was taken as $R_d =
H_z(\frac{t-t_{conf}}{\tau(E)})$
and the term $Q = 1$ at $r < R_s$ within the shell and 
$Q = (1+(\frac{r-R_s}{R_d})^2)^2$ at $r > R_s$ outside the
shell. Other parameters were explained before, {\em R} is taken as 288 pc.
The term $Q(E,r,t)$ determines the {\em shape} of the LDF and the volume (9) 
arises as the integral 
$V_{cr}=\int_0^{\infty} 4 \pi r^2 Q(E,r,t) dr$
which has a meaning of the effective volume occupied by cosmic rays. It has been used for
the determination of the cosmic ray concentration in the central part of the 
SNR, i.e. at $r < R_s$.
\begin{figure}[htbp]
\begin{center}
\includegraphics[width=15cm,height=15cm,angle=0]{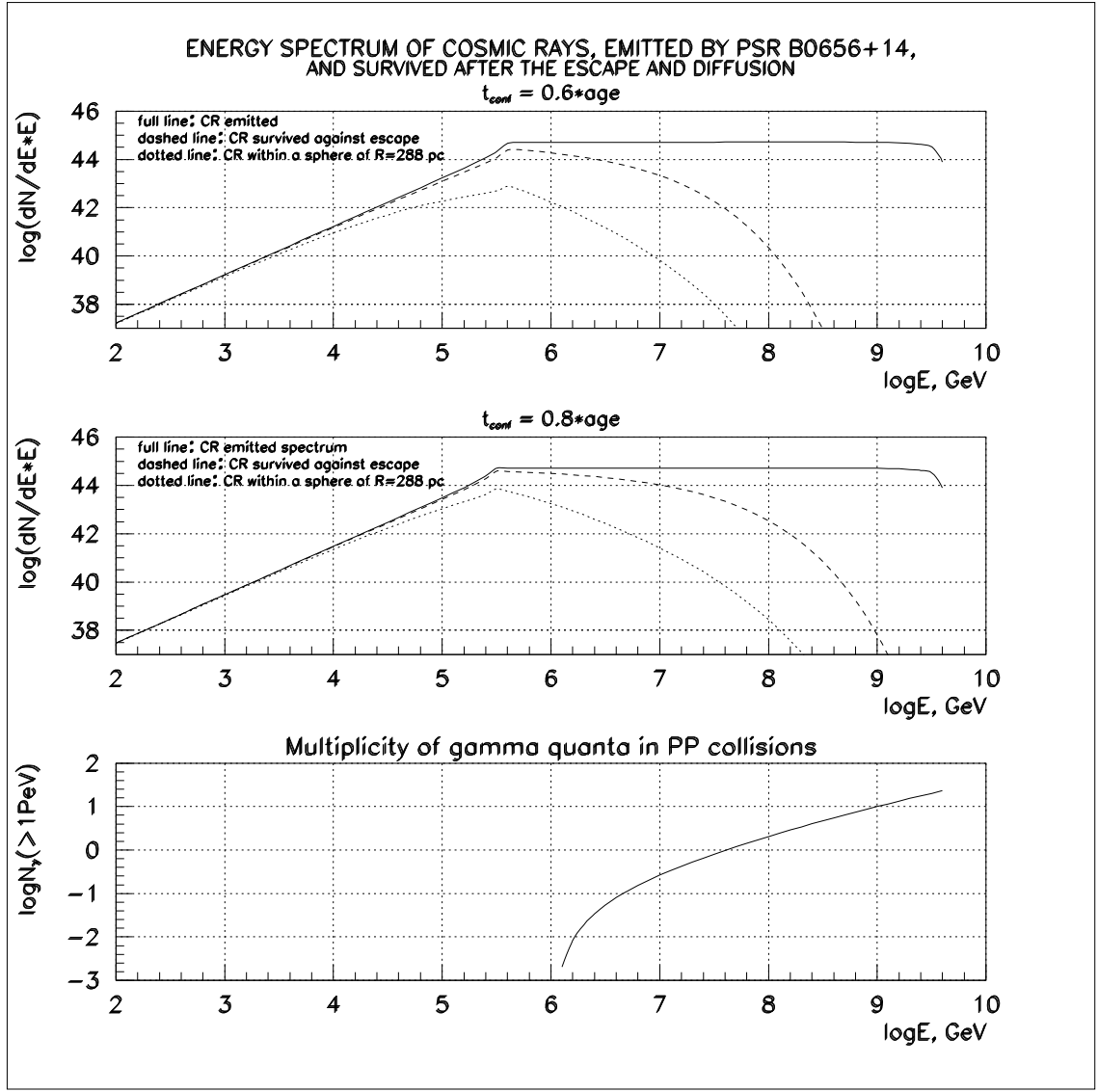}
\caption{\footnotesize Energy spectrum of cosmic rays emitted by pulsar B0656+14 during
its life time and survived after the escape and diffusion: upper panel - for 
$\Delta=0.6$, middle panel - for $\Delta = 0.8$. Full line:
the emitted spectrum, dashed line - the same spectrum surviving against escape from 
the Galaxy, dotted line - the spectrum of cosmic rays remaininig around the pulsar within
the radius R after escape and diffusion. The lower panel shows multiplicity of 
gamma quanta with energy above 1 PeV in PP collisions as a function of the proton
energy.}
\end{center}
\label{fig:puls4}
\end{figure}

Calculation of the flux for $\Delta = 0.8$ (~dotted line in the middle 
panel of Figure~4~) gives a value of $3.7\cdot 10^{-15} cm^{-2}s^{-1}$,
which is less than the experimental value $\sim10^{-13}
cm^{-2}s^{-1}$ by more than an order of magnitude. However, the
calculations show a strong dependence of the gamma ray flux on the value
for the
time when the cosmic rays were released from confinement. For
instance, if $\Delta = 0.95$ the flux rises up to $2.0\cdot
10^{-13} cm^{-2}s^{-1}$ and exceeds the experimental value by the
factor of $\sim 2$. For  $\Delta = 0.99$ the flux is
 $2.2\cdot 10^{-12} cm^{-2}s^{-1}$, which is by an order of magnitude
higher than the previous value. 

If the pulsar accelerates nuclei, then their energy at the fixed
rigidity will be $Z$ times higher and they are more efficient in the
production of gamma-quanta \cite{EW3}. However, because the total
energy lost by the pulsar for the acceleration of particles is fixed,
the spectrum of nuclei at the same energy is $Z$ times lower than that
of protons. Since the mean production rate for gamma quanta by nuclei
of the mass $A$, which is proportional to the product of the cross
section and the number of wounded nucleons, is $A$ times higher, then
the spectra of gamma quanta from interactions of nuclei would be 
$A/Z \approx 2$ times higher than that for protons.

\subsubsection{Electron interactions}

As for mechanisms of direct production of PeV gamma rays the
electromagnetic interactions of accelerated electrons can be
discussed. The energy spectrum of electrons accelerated by the pulsar is 
similar to the rigidity  spectrum of nuclei, since all the particles
are accelerated by the electric field with the same potential
difference. 

Among the direct processes, the inverse compton scattering on microwave 
background photons can be excluded, since in the energy spectrum of
electrons there are no energies which can boost the background
photons with a mean energy of $\sim 10^{-4} eV$ up to the PeV energies.

The mechanism which {\em might} be possible from the energy requirements
and interaction kinematics is the inverse compton scattering of
accelerated electrons on X-ray photons, The Monogem Ring is
known as a source of X-rays \cite{Bunne,Nouse,Pluci} and inverse
compton scattering of electrons on them might produce the needed PeV 
gamma-quanta. However, the problem with this mechanism is
that electrons which are able to boost keV X-rays up to PeV energies 
(~$> 5\cdot 10^8$ GeV~) are produced by B0656+14 pulsar only during the
first few decades. We have to confine them without substantial
energy losses during $\sim 10^5$ years up to the present time to let
them scatter on X-ray photons and give the observed PeV gamma quanta.
Such a long confinement time for EeV electrons without any substantial
energy losses seems to be unlikely. Moreover, even if such a
confinement is possible the intensity of the produced gamma quanta is
by about 14 orders of magnitude less than that from PP-interactions
( 7 orders of magnitude are due to the difference in the cross sections
and another 7 orders - due to the low density of X-ray photons ).

Therefore electrons could not be the source of the gamma-quanta in the
Armenian peak.    
      
\subsubsection{Neutrons}

Neutrons being able to travel in straight lines can give rise to the narrow peak of
EAS intensity from the discrete source. The processes which produce
neutrons are PP-interactions with charge exchange for the projectile 
proton and disintegration of heavier nuclei, if they are emitted by
the pulsar. The minimum energy of neutrons which can give the observed EAS 
with $N_e > 10^6$ is higher than for gamma-quanta and for the Aragats
altitude is equal to 2.5 PeV. We calculated the flux of neutrons from
PP-interactions similarly to that of gamma quanta:
\begin{equation}
F_n(>3PeV) = \int_{2.5PeV}^{E_{max}^0}dE_n\int_0^R D(E_n,r)
dr\int_{2.5PeV}^{E_{max}^0} \frac{dN}{dE} \rho_{cr} \sigma_{in} n_n(E,E_n) \rho_{ISM} dE 
\end{equation}
Here most notations are the same as in (8) and $E_n$ is
the neutron energy, $n_n(E,E_n)$ is the inclusive spectrum of
neutrons produced in PP-collisions and $D(E_n,r)$ is the survival
probability of neutrons. decaying on their way from the birth
place to the Earth. We have taken the inclusive spectrum of neutrons,
$n_n(E,E_n)$, from the experimental data taken at 24 GeV \cite{Blobe}
and asssumed that they can be scaled up to PeV energies in terms of
their $x = E_n/E$ dependence. The survival probability has been taken
as
\begin{equation}
D(E_n,r)=exp(-\frac{R-r}{l_d(E_n)})+exp(-\frac{R+r}{l_d(E_n)})
\end{equation}
with $l_d(E_n)$ as the decay length of neutrons. 

The influence of the higher energy threshold, which leads to a
 smaller flux of the relevant protons and their faster diffusion
 from the confinement volume with a further reduction of their
 density, lower multiplicity of neutrons produced in PP-interactions
 and their decay results in a substantially lower flux
 of neutrons compared with that of gamma-quanta. For $\Delta =
 0.8$ the flux $F_n(>2.5PeV) = 4.0\cdot 10^{-20} cm^{-2}s^{-1}$, for
 $\Delta = 0.95$ - $7.4\cdot 10^{-18} cm^{-2}s^{-1}$. 

A better opportunity is provided by heavier nuclei, if they are indeed 
accelerated by the pulsar. They have a higher cross-section for 
interaction with protons of the ISM and release multiple spectator neutrons
with the same energy per nucleon without energy losses. However, the
gain is not enough because of the need for energy conservation. Since the
energy lost by the pulsar for the acceleration of nuclei is fixed, the
number of nuclei with the same rigidity is respectively less. Moreover,
to get neutrons with $E_n > 2.5 PeV$, pulsars have to accelerate nuclei
to higher rigidity than protons in order to get the same energy
per nucleon. All these factors lead to fluxes of neutrons higher
only by a factor less than 10. For example, if the pulsar
accelerates iron nuclei and they release all their 30 neutrons after
collision with ISM protons, then for $\Delta = 0.8$, the flux
$F_n(>2.5PeV) = 1.2\cdot 10^{-19} cm^{-2}s^{-1}$, for $\Delta = 0.95$
it is $3.7\cdot 10^{-17} cm^{-2}s^{-1}$.

Therefore, neutrons, too, cannot be the particles which give rise to the
Armenian peak.    

\section{Discussion}

We conclude that although the pulsar B0656+14 cannot be the domimant source
supplying cosmic rays in the knee region, it alone accelerates protons
which can produce gamma quanta observable as an excess EAS intensity
in the Armenian peak. Certainly cosmic rays from the associated SNR
Monogem Ring can contribute to this intensity. The only condition is
that cosmic rays from this pulsar should be confined by the associated
SNR during a considerable fraction of its age.
  
\subsection{Age of the pulsar and the duration of the confinement}

The spin-down age of the pulsar B0656+14, determined as $age =
P/2\dot{P}$, is equal to $1.1\cdot 10^5$ year. However, if indeed
the higher multipoles of the magnetic fiels \cite{Phinn,EW10} or 
gravitational radiation \cite{Ostri} are important, the actual energy
losses at the initial period of the pulsar evolution can be higher and
its observed rotation parameters could be reached in a shorter time,
leading to a smaller age. The model age is estimated as $8.6\cdot
10^4$ year \cite{Pluci}. If we use this shorter age the estimated 
flux of gamma quanta for $\Delta = 0.95$ rises up to 
$2.9\cdot 10^{-13} cm^{-2}s^{-1}$, leaving a safe gap of $<40\%$
efficiency for the conversion of the pulsar energy into cosmic rays.
The fraction of $\sim$95\% of the age, required for the confinement time
means that the cosmic rays were trapped inside the SNR for about 82 kyear.
Remarkably, this value coincides with the 80 kyear adopted by our
model of the SN
explosion and acceleration of cosmic rays \cite{EW1}. It means that
cosmic rays had only a few kyears to diffuse and that is why only high
energy particles reached the Earth forming the sharp knee.

\subsection{The contribution of B0656+14 beyond the knee}

If the age is fixed and the confinement time increases, less time
remains for diffusion and escape. The observed spectrum of cosmic
rays then approaches the emitted spectrum. Since the emitted spectrum is
as flat as $E^{-1}$ \cite{Ostri,Blasi,Gille} then if the fraction
$\Delta$ approaches 0.9, then the intensity of cosmic rays from the
pulsar B0656+14 calculated in \$4.4 exceeds the observed values at
high energies beyond the knee. In
principle it might create problems for our explanation of the origin of the
Armenian peak. However, we remember that calculations were made
for 100\% efficiency of the conversion of the pulsar rotation
energy into cosmic rays and for a {\em mean} ISM density $\rho_{ISM} = 3\cdot
10^{-3} cm^{-3}$. Since we assume that the Armenian peak is due to 
interactions of cosmic rays with the local perturbation of the ISM
density we can safely reduce the efficiency of conversion not to
exceed the observed cosmic ray intensity and increase
the local ISM density respectively to preserve the same flux of
gamma-quanta as observed in the peak.

Our estimate of the flux has been made as the integral along the
line of sight through the whole SNR. However, the size of the bin for the
Armenian peak is $\sim$10 times smaller than the size of the SNR. In
order to get the same grammage, for which we obtained the flux
comparable with observed, we have to increase the gas density in
the local perturbation by the factor of $\sim$10 up to $\sim 3\cdot
10^{-2} cm^{-3}$. Since the mean gas density in the Galaxy is $\sim 1
cm^{-3}$ such an increase seems possible.

We should also remark that the contribution of the pulsar
B0656+14 to cosmic rays beyond the knee might be considerable and grow
with energy up to $10^8$ GeV. If it
is true, and taking into account that the mass composition of cosmic
rays beyond the knee becomes heavier with a growing amount of
iron \cite{Haung,Hoera}, we have to conclude that the pulsar B0656+14
can emit iron nuclei. 

\subsection{The connection between the pulsar and the SNR}

The confinement of high energy cosmic rays by the SNR raises other
intersting problems: how can the Monogem Ring SNR, which according to
our model accelerates cosmic rays only up to 0.4 PV, rigidity trap and
confine for a long time particles with rigidities up to $10^3$ PV ?
A possible idea for such a scenario is that SNR with their
large sizes and moderate magnetic fields are efficient in the
acceleration of relatively large fluxes of particles up to moderate
sub-PeV and PeV energies, whereas a pulsar which has much higher
magnetic fields in a much smaller volume, can accelerate smaller 
fluxes of particles reduced further by beaming, but up to much higher 
super-PeV energies. It means
again that the pulsar B0656+14 and, probably, other pulsars, might be
serious contenders for the sources of cosmic ray particles beyond the knee,
discussed often as the so called `second component' of cosmic rays in two-
or three-component models (~eg. \cite{Wdowc,Ficht,Gaiss,Danil,Teran}).

\subsection{Position of the Armenian peak and the pulsar B0656+14}

There is a potential difficulty in the association of the
Armenian peak with the pulsar. It is the mutual position of the region
where the EAS intensity peak is observed , the pulsar position and the
direction of its proper motion. Chilingarian et al. remark that
despite the fact that their intensity peak is inside the Monogem Ring
SNR it is displaced from the pulsar position by 8.5$^\circ$ (~Figure
5~). It is rather far. Moreover, if the excess EAS intensity is due to
the interaction of cosmic rays produced by the pulsar in the past,
then the direction of its proper motion should be away from the
region where it was in the past. On the contrary, the direction of
B0656+14's proper motion is towards the region of the peak. 
\begin{figure}[htb]
\begin{center}
\includegraphics[width=11cm,height=9cm,angle=0]{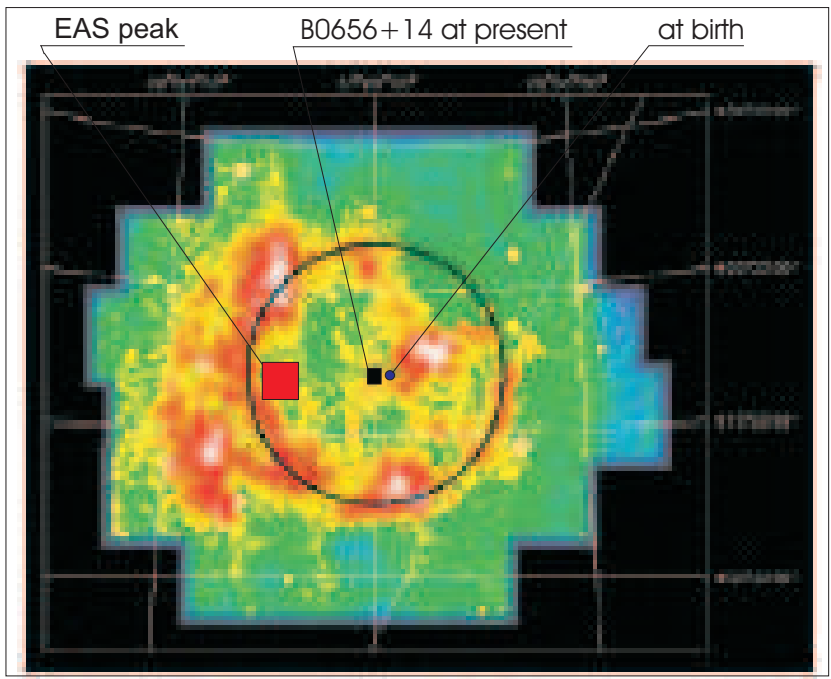}
\caption{\footnotesize The Monogem Ring, as seen in the ROSAT all-sky survey in the 
0.25-0.75 keV X-ray band. PSR is at the center of the Figure and the
circle is of the radius 9.2$^\circ$ centered 
on this point. The circle indicates the primary ring structure. The position of the pulsar 
10$^5$ years ago, estimated from its proper motion, is marked with a
small square about 1$^\circ$ to the right of its present position
(~this figure was copied from \cite{Thors}~).  
The angular bin where Chilingarian et al. have found an excessive intensity of EAS 
is shown by the big square, marked 'EAS peak'} 
\end{center}
\label{fig:puls5}
\end{figure}

It must be remarked that there may be a problem with the absolute
celestial alignment of the MAKET-ANI array \cite{Chil2} and this
problem may go away. 

However, even if everything is correct, our scenario with a long 
confinement could give an explanation
for the possible misalignment of the pulsar and the Armenian peak. 
The diffusion radius of PeV protons reached in
4 kyears is 100 pc and a sphere of this radius can be seen from a  
distance of 300 pc at an angle of about 20$^\circ$. It means that PeV
cosmic rays have overcome the parent pulsar and its proper motion is
not necessarily connected with the regions of the highest cosmic ray
density. Within the sphere the density of cosmic rays is presumably 
uniform and any kind of local
ISM density perturbation or a local molecular cloud could create the
excess intensity of gamma quanta. In this connection it would be
interesting to search for other possible excesses in this area.        

\subsection{The search for other EAS intensity excesses in MAKET-ANI data}

The results from the experiment of Chilingarian et al. \cite{Chil1}
are so potentially important that further analysis is clearly worthwhile.
The question to be asked is: irrespective of the likely association of
the main peak with the pulsar in the Monogem Ring, is the rest of the data
consistent with there being a whole family of sources in the PeV
region ? This possibility is not fanciful: there is the well-known
presence of `unidentified' GeV sources and at least one TeV peak
\cite{Aharo} does not (~yet~) appear to have been identified. 

In terms of pulsars of age less than $10^5$y we expect $\sim$1000 in
the Galaxy at the present time (~$10^{-2}y^{-1}$ birth rate~). The MAKET-ANI
array covers about 30\% of the sky and about 50\% of the Galactic
Plane region (~$|b|\simeq 20^\circ$~). Thus $\sim$500 sources might be
present. Using the analysis already described, those within $\sim$0.5
kpc might be directly discernible, viz $(0.5/15)^2 \times 500, \simeq$
0.5. The potential Monogem source falls in this category. The question
is the presence, or otherwise, of weaker sources and their
contribution to the overall flux.

Inspection of Figure 6(a), taken from \cite{Chil1}, which relates to
distribution of significances for the whole sky seen by the array,
shows evidence for an excess of about 11 `events' beyond $\sigma \sim
2$. The corresponding fluxes are $>$1/9 times the Armenian peak flux.
If all the events were along the Galactic Plane we would have expected
of order 0.5$\times$9 i.e. $\simeq$5. In fact, the Galactic latitudes of
the events in Figure 6(a) are not known but it is evident that a good
case can be made for some extra, weak sources. Indeed, if, following
our arguments \cite{EW10} for spun up pulsars, at high Galactic
latitudes, being responsible for some of the very energetic particles,
we could imagine a wider distribution in Galactic latitude. The
expected number would be correspondingly greater (~for 3-D compared
with 2-D, the number would go up by $9^{\frac{3}{2}}$ = 27 cf 9~).

Figure 6(a) is reminiscent of the result from the Tibet-III Air Shower
Array \cite{Cui}. In a similar significance plot, upward deviation
occur increasingly above 3$\sigma$, reaching a factor 2 above 
4$\sigma$.

Figure 6(b) is our own plot of the array data in terms of deviation
from the mean for the restricted declination range 
(~$\delta: 12.5^\circ-15.5^\circ$~) which crosses Monogem.  Again, we
see evidence for a small excess of positive excursions from the mean
for positive values. Examining the basic data for this $\delta$-range,
we find no evidence for an excess, of large fluxes near the Galactic
Plane, however. Again, the sources may be more widely distribute than
just in the Galactic Plane.
\begin{figure}[htbp]
\begin{center}
\includegraphics[width=15cm,height=6cm,angle=0]{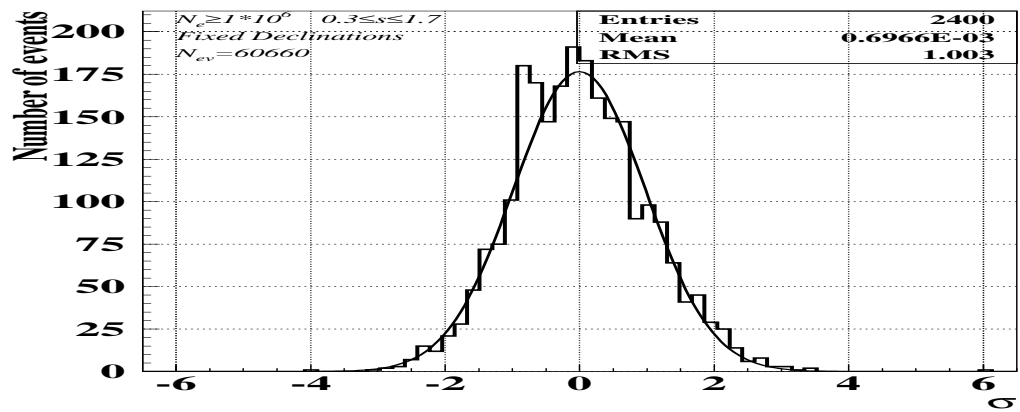}
\includegraphics[width=15cm,height=6cm,angle=0]{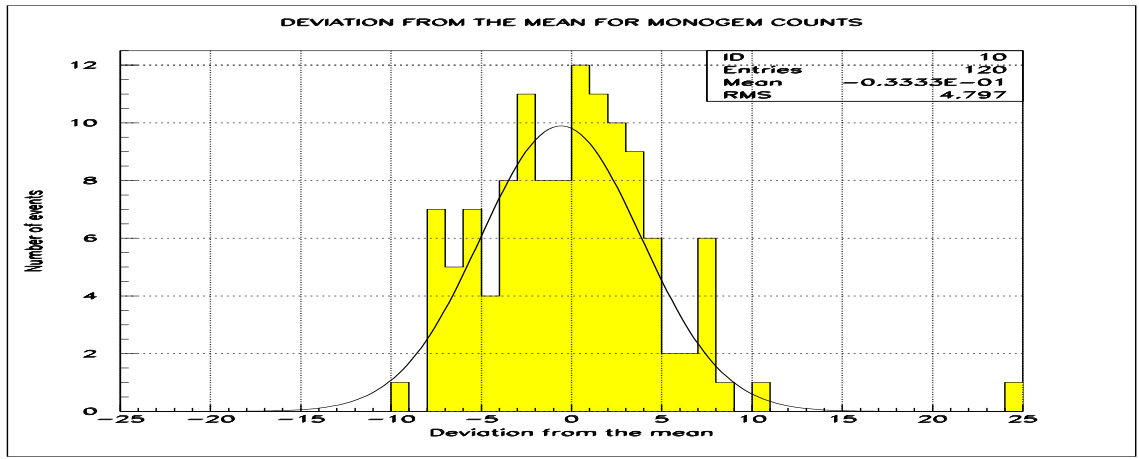}
\caption{\footnotesize (a) Signal significance test with full
equatorial coverage with 2400 $3^\circ \times 3^\circ$ bins; 
$N_e > 10^6$. (b) Frequency distribution of deviations from the
mean in the number of events per 3$^\circ$RA bin in the cut along
$\delta: 12.5^\circ - 15.5^\circ$. The data relate to the EAS
observations of Chilingarian et al. \cite{Chil1}}
\end{center}
\label{fig:puls6}
\end{figure}

\subsection{Possible contribution of other pulsars to the knee}

It is well known that some pulsars emit energetic gamma rays (~see
\cite{Weeke} for a review~). If we regard the early `observation' of Cygnus
X-3 \cite{Samor,Watso} as showing that pulsars {\em do} emit PeV gamma
rays, albeit spasmodically, then a way forward appears. Cyg X-3
appears to have given a flux above 1 PeV of 
$\approx 3 \cdot 10^{-14} cm^{-2}s^{-1}$ when `{\em on}'; its distance
is 11.4 kpc, so
that fluxes of order $4\cdot 10^{-11} cm^{-2}s^{-1}$ would be expected
for sources at $\sim$300 pc. There would therefore be no problem of flux for
B0656+14, with its measured flux of $10^{-13} cm^{-2}s^{-1}$. Even the
many upper limits from other workers who have searched for Cygnus X-3,
and which are typically a factor 100 below the `observations', would
not be inconsistent with the Armenian peak expectation.

Bhadra examined the possibility for pulsars to form the knee and
concluded that Geminga and Vela are the most likely candidates
\cite{Bhadr}. We calculated energy spectra  of cosmic rays produced by
these pulsars, considering them as isolated pulsars; 
they are shown in Figure 7.

Since Geminga is older than
B0656+14 (~its age is $3\cdot 10^5$ y~), its maximum rigidity will be 0.19
PV, therefore it could contribute to the knee energies only if it
accelerates iron nuclei. In this case there is no room for the `second
knee' in the interval 10-20 PeV found by us and included in the
Single Source Model as an 'iron peak'.     

Vela is much younger (~$\sim 10^4$ year~). Its maximum rigidity is
right in the knee region of $\sim$2.8 PV, so that it could contribute 
even by accelerating protons. The intensity can be adjusted
by introducing the efficiency of conversion of the pulsar rotation
energy to the accelerated particles at the level of 10\%, which is
a reasonable value. However, the Vela pulsar is associated with a young
SNR and most likely its accelerated  cosmic rays are still confined in the
SNR shell.  
\begin{figure}[htp]
\begin{center}
\includegraphics[width=9cm,height=9cm,angle=0]{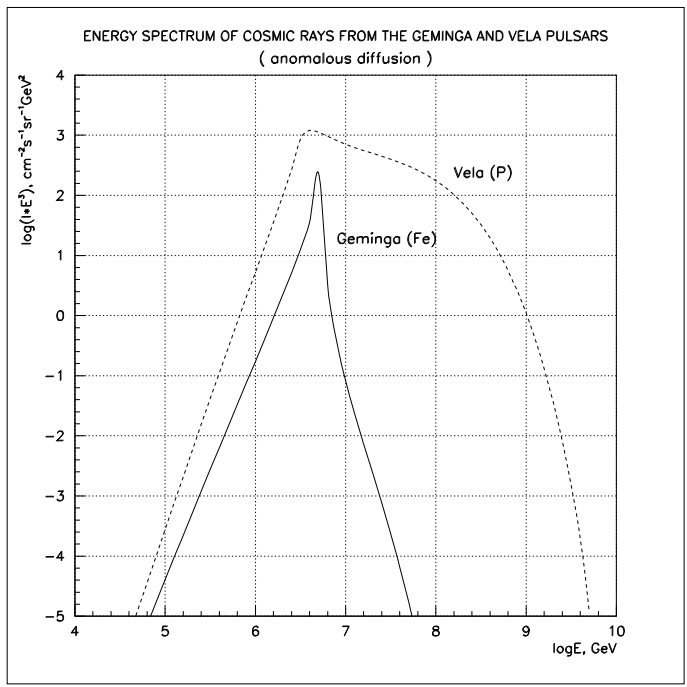}
\caption{\footnotesize Energy spectra of cosmic rays produced by
the Geminga and Vela pulsars. Full line: Geminga, accelerating iron nuclei,
dashed line: Vela, accelerating protons.}
\end{center}
\label{fig:puls7}
\end{figure}

\subsection{Predictions}

The scenario described here let us make three predictions for the Monogem
Ring results, which could be checked experimentally:

(i) since the only mechanism responsible for the formation of the EAS
intensity peak appears to be gamma ray production in PP-collisions, the excess
showers in the peak should have different characteristics from the
background showers. They should have features typical of
electromagnetic cascades ;

(ii) even if the peak is created by the pulsar, the fact that cosmic
rays responsible for it were produced in the past, were confined and
most likely mixed with cosmic rays from the SNR, the gamma ray
emission should not be pulsed ;

(iii) due to the non-uniform character of the ISM there might be some
other excesses in the Monogem Ring area due to other local
perturbations of the ISM density, most likely of smaller amplitude.   

\section{Conclusions}

We conclude that the pulsar B0656+14 can contribute to the intensity of cosmic rays
in the knee region, but cannot be the dominant source responsible for its formation. 
Its contribution to the intensity of the Single Source, needed to form the sharp knee,
appears not to exceed 15\%.  
The SNR associated with the Monogem Ring, rather than the pulsar, still remains the most likely Single Source 
which gives the dominant contribution to the formation of the cosmic ray energy 
spectrum in the vicinity of the knee.    

We have also examined the possibility of the pulsar B0656+14 giving
the peak (~the`Armenian peak'~) of the EAS intensity, observed from
the region inside the Monogem Ring. The estimates of the
gamma-ray flux produced by cosmic ray protons from this pulsar
evidence that it can be the source of the observed peak, if the
protons were confined within the SNR during a considerable fraction
(~$\sim$90\%~)of its
total age. The flux of gamma quanta at PeV energies has a high
sensitivity to the duration of the confinement. The estimate of this 
time and the following diffusion of cosmic rays from the confinement
volume turns out to be
in remarkable agreement with the time needed for these cosmic rays to
propagate to the solar system and to form the observed knee in the
cosmic ray energy spectrum. 

Other possible mechanisms for the production of particles which could
give rise to the Armenian peak were also
examined. Electrons scattered on the microwave
background or on X-rays, emitted by SNR, could not be responsible for
the gamma-quanta in the peak. 
Neutrons produced in PP - collisions or released from the
spallation of accelerated nuclei also seem to be improbable mechanisms 
since they cannot give the observed flux.  

    If the EAS results are confirmed, they will be important, since\\
(i) they give evidence for the possibility of the acceleration of protons by
the pulsar; \\
(ii) they give evidence for the existence of a confinement mechanism in SNR
;\\
(iii) they confirm that cosmic rays produced by the Monogem Ring SNR
and associated pulsar B0656+14 were released recently giving rise to
the formation of the sharp knee and the observed narrow peak in the
EAS intensity ;\\
(iv) they give strong support for the Monogem Ring being
identified as the source proposed in our Single Source Model
of the knee.

\vspace{0.5cm}

{\large{\bf Acknowledgements}}

Authors thank A.A.Chilingarian for useful discussions and an unknown
referee for the remarks. One of the authors (ADE) thanks The Royal
Society for financial support.

\end{document}